\documentclass[cameraready]{Interspeech}

\usepackage{amsmath}
\usepackage{booktabs}
\usepackage{siunitx}
\usepackage{graphicx}
\usepackage{url}
\usepackage{hyperref}
 \usepackage{float}
\usepackage[most]{tcolorbox} 
\usepackage{xcolor}          
\usepackage{subcaption}    
\title{Quantifying Cross-Lingual Transfer in Paralinguistic Speech Tasks}

\author[affiliation={1,2}, orcid=0009-0005-1055-129X, correspondingauthor]{Pol}{Buitrago}
\author[affiliation={1}, orcid=0009-0001-4084-1598]{Oriol}{Pareras}
\author[affiliation={1,2}, orcid=0000-0002-1389-3595]{Federico}{Costa} 
\author[affiliation={1,2}, orcid=0000-0002-1730-8154]{Javier}{Hernando}

\address{
    $^1$ Barcelona Supercomputing Center (BSC), Spain\\
    $^2$ Universitat Politècnica de Catalunya (UPC), Spain
}

\email{pol.buitrago@bsc.es}


\begin{document}

\maketitle

\begin{abstract}

Paralinguistic speech tasks are often considered relatively language-agnostic, as they rely on extralinguistic acoustic cues rather than lexical content. However, prior studies report performance degradation under cross-lingual conditions, indicating non-negligible language dependence. Still, these studies typically focus on isolated language pairs or task-specific settings, limiting comparability and preventing a systematic assessment of task-level language dependence. 

We introduce the Cross-Lingual Transfer Matrix (CLTM), a systematic method to quantify cross-lingual interactions between pairs of languages within a given task. We apply the CLTM to two paralinguistic tasks, gender identification and speaker verification, using a multilingual HuBERT-based encoder, to analyze how donor-language data affects target-language performance during fine-tuning. Our results reveal distinct transfer patterns across tasks and languages, reflecting systematic, language-dependent effects.

\end{abstract}

\section{Introduction}

Multilingual training and cross-lingual transfer are widely used strategies to leverage knowledge from resource-rich languages to improve performance on languages with limited labeled data \cite{ji-etal-2024-machine,pires-etal-2019-multilingual,EasyChair:12273,SINDANE2025100157,manafi-krishnaswamy-2024-cross}. Previous studies have shown that adding training data from other languages can either improve or degrade the performance of a target language \cite{wang2020negativeinterferencemultilingualmodels,pires-etal-2019-multilingual,EasyChair:12273,SINDANE2025100157}, yet the patterns governing when cross-lingual transfer is beneficial remain unclear. Recent advances in multilingual self-supervised speech models suggest that large-scale training can produce representations capturing cross-lingual acoustic–phonetic structure \cite{babu2021xlsrselfsupervisedcrosslingualspeech,baevski2020wav2vec20frameworkselfsupervised,hsu2021hubertselfsupervisedspeechrepresentation}, but it remains unclear how to quantify the impact of one language’s data on another’s performance.

In speech, this problem is intrinsically entangled, as linguistic content, prosody, and paralinguistic cues are tightly intertwined \cite{SCHULLER20134,lin2020prosody}. Tasks such as automatic speech recognition or machine translation rely primarily on lexical and semantic information, whereas tasks based on extralinguistic cues, like gender identification or speaker verification, show less clear language dependence. At first glance, these traits appear intrinsic to the individual and largely independent of linguistic content. However, prior work has shown that performance on paralinguistic tasks can still degrade significantly under language mismatch \cite{1327105,940862,DBLP:conf/slt/MisraH14,7919004}.

Existing studies on cross-lingual transfer in paralinguistic tasks have mostly focused on a few language pairs and task-specific setups \cite{gender_anidjar,li2017crosslingualspeakerverificationdeep,saad2022casestudyindependencespeech,Siddique2017,Xia_2019}, making it difficult to draw systematic conclusions about task-level language dependencies.

Several approaches have been proposed to quantify cross-lingual transfer, each addressing different aspects of this complexity. Simple metrics, such as subword overlap \cite{deshpande2022bertmultilingualisolatingcrucial} or alignment of representations across parallel sentences \cite{kargaran2025mexamultilingualevaluationenglishcentric}, provide predictive signals of transfer without an explicit matrix formulation, but do not capture task-level donor–target interactions. 

A related line of work evaluates cross-lingual transfer via single-source adaptation. In \cite{dymkiewicz2026donorsrecipientsasymmetrictransfer}, a model is fine-tuned on a single language pair and absolute performance changes are measured on other languages, while \cite{blaschke2025analyzingeffectlinguisticsimilarity} constructs zero-shot matrices by training on one language and evaluating on others. Both approaches use only monolingual data and do not normalize performance for comparability across languages.

In contrast, other methods construct explicit transfer matrices, such as the Interference Matrix \cite{alastruey2025interferencematrixquantifyingcrosslingual}, which measures how joint bilingual training affects each language’s loss; or ATLAS \cite{longpre2025atlasadaptivetransferscaling}, which reports large-scale pair-wise cross-lingual gains in multilingual pretraining. While these approaches capture interactions between languages, they are not directly grounded in downstream performance, nor suitable for comparing architectures or task-specific setups under a common framework.

Existing methods either focus on alignment of model representations, absolute gains in single-source adaptation, or large-scale pretraining improvements, but none offer a framework to quantify donor effects on downstream performance, enabling systematic comparison across heterogeneous tasks. To address this gap, we propose the Cross-Lingual Transfer Matrix (CLTM), a normalized pairwise measure of cross-lingual transfer that captures the change in downstream performance induced by adding donor-language data during fine-tuning. The CLTM offers a performance-based evaluation framework, with metrics quantifying donor effects on target performance, enabling systematic comparison across tasks and architectures.

To validate the proposal, we apply the CLTM to two representative paralinguistic downstream tasks, Gender Recognition (GR) and Speaker Verification (SV), using a rigorously controlled experimental protocol across 44 languages. All experiments are conducted using the same multilingual pretrained HuBERT encoder, which is adapted to each task through consistent fine-tuning procedures. By fixing the underlying architecture and controlling initialization, data regimes, and training conditions, we try to isolate donor-language effects and systematically evaluate cross-lingual effects under consistent conditions.

The aim of this work is twofold: first, to define and validate the CLTM as a performance-grounded measure of cross-lingual transfer; and second, to apply it to two example paralinguistic tasks, characterizing patterns of inter-language transfer.

\section{Method}
\label{sec:method}

This section introduces the CLTM and the framework required to compute it. We define the CLTM, describe how it quantifies the impact of donor-language data on target-language performance, and outline the validity conditions needed to ensure comparable transfer estimates across language pairs.

\subsection{Cross-Lingual Transfer Matrix (CLTM)}
\label{sec:cltm}

Let $D_l$ and $D_l'$ be two non-overlapping sets of training data from language $l$, each containing $N$ samples.
We define $\mathrm{Perf}_l(D_l)$ as the performance of a model trained on $D_l$ and evaluated on language $l$, measured with a higher-is-better metric.

Given a target language $i$ and a donor language $j$, we define the \emph{self-gain} and the \emph{cross-gain} as
\begin{equation}
\begin{aligned}
\Delta_{i\leftarrow i} &= \mathrm{Perf}_i(D_i + D_i') - \mathrm{Perf}_i(D_i),\\
\Delta_{i\leftarrow j} &= \mathrm{Perf}_i(D_i + D_j) - \mathrm{Perf}_i(D_i).
\end{aligned}
\end{equation}

The Cross-Lingual Transfer Matrix (CLTM), therefore, is the row-normalized matrix
\begin{equation}
\mathrm{CLTM}[i,j]
=
\frac{\Delta_{i\leftarrow j}}{\Delta_{i\leftarrow i}},
\label{eq:cltm_def}
\end{equation}\vspace{-0.25cm}
defined for $\Delta_{i\leftarrow i}>0$, and satisfying $\mathrm{CLTM}[i,i]=1,\ \forall i$.\\ 

\begin{tcolorbox}[colback=white, colframe=black!50, boxrule=0.5pt, arc=2mm, left=2mm, right=2mm, top=1mm, bottom=1mm]
\textbf{Interpretation.} Each CLTM entry indicates how donor-language $j$ affects target language $i$ relative to an equivalent amount of target-language data:
\begin{itemize}
  \item $\mathrm{CLTM}[i,j] < 0$: donor data decreases performance.
  \item $0 < \mathrm{CLTM}[i,j] < 1$: donor-language data improves performance, but less than target-language data.
  \item $\mathrm{CLTM}[i,j] > 1$: donor data improves performance more than the same amount of target-language data.
\end{itemize}
\end{tcolorbox}\vspace{0.1cm}

The CLTM is a row-normalized matrix, which ensures direct comparability across different languages. The diagonal is fixed at one and off-diagonal entries quantify donor effects. 

For a fully language-agnostic task, all entries would equal one ($CLTM=\mathbf{1}_{n\times n}$), indicating that all donors help all targets equally. To systematically characterize cross-lingual transfer, we define the following metrics:\vspace{0.2cm}

\noindent\textbf{Relative Frobenius Deviation (RFD)}\;~Quantifies how far the CLTM deviates from the ideal agnosticity matrix $\mathbf{1}_{n\times n}$. Small values indicate a nearly language-independent task; large values indicate strong language-specific effects:
\begin{equation}
\mathrm{RFD_1} = \frac{\lVert \mathrm{CLTM} - \mathbf{1}_{n\times n} \rVert_F}{\lVert \mathbf{1}_{n\times n} \rVert_F} = \frac{\lVert \mathrm{CLTM} - 1 \rVert_F}{n}.
\end{equation}

\noindent\textbf{Relative Asymmetry}\;~Captures differences in transfer when the roles of donor and target are reversed. Zero denotes symmetry, higher values denote directional bias:
\begin{equation}
\mathrm{Asym}_{\mathrm{rel}} = \frac{\lVert \mathrm{CLTM} - \mathrm{CLTM}^\top \rVert_F}{\lVert \mathrm{CLTM} \rVert_F}.
\end{equation}

\noindent\textbf{Average Row Cosine Similarity}\;~Captures the similarity of transfer profiles across target languages. High values indicate that different targets benefit from donors in a similar manner:
\begin{equation}
\overline{\cos}_{\mathrm{rows}} = \frac{1}{n(n-1)} \sum_{i=1}^{n} \sum_{\substack{j=1 \\ j\neq i}}^{n} 
\frac{\mathrm{CLTM}[i,:]\cdot \mathrm{CLTM}[j,:]}{\lVert \mathrm{CLTM}[i,:] \rVert \, \lVert \mathrm{CLTM}[j,:] \rVert}.
\end{equation}

We also define complementary statistics derived from the off-diagonal entries, namely the proportion of positive transfer (prop$_+$), the proportion of reciprocal positive interactions (reciprocity$_+$), and the proportion of positive transfer within language families (intra-family$_+$).

\subsection{Dynamic Training Interval}
\label{sec:dat_regime}

The CLTM is meaningful only if performance changes from additional data are reliably measurable. We select a task-specific training interval $[N,2N]$ where the model is neither undertrained nor near performance saturation. 

To choose this range, we perform preliminary experiments for each language, inspecting the learning curves to identify the region where performance grows substantially with additional data (the ``Dynamic'' region in Figure~\ref{fig:learning_curve_clean}). If $N$ is too small, language-specific effects cannot be distinguished; if too large, performance plateaus and donor-language data has little impact. The selected intervals ensure that $\mathrm{Perf}_i(D_i)$ and $\mathrm{Perf}_i(D_i + D_i')$ fall within this dynamic regime for all experiments. \vspace{-0.1cm}

\begin{figure}[h!]
\centering
\resizebox{0.95\linewidth}{!}{%
\begin{tikzpicture}[x=2.2cm,y=1cm, every node/.style={font=\LARGE}]

  \pgfmathsetmacro{\xmin}{0}
  \pgfmathsetmacro{\xmax}{8.2}   
  \pgfmathsetmacro{\ymin}{0}
  \pgfmathsetmacro{\ymax}{6}

  \draw[->, line width=2pt] (\xmin,\ymin) -- (\xmax,\ymin)
    node[below=14pt, left=2pt] {Data volume}; 
  \draw[->, line width=2pt] (\xmin,\ymin) -- (\xmin,\ymax+0.25)
    node[left=18pt, rotate=90] {Performance}; 
    
  \fill[red!15] (0,\ymin) rectangle (2.5,\ymax);        
  \fill[green!15] (2.5,\ymin) rectangle (5.0,\ymax);    
  \fill[yellow!15] (5.0,\ymin) rectangle (8.0,\ymax);   

  \draw[very thick, line width=2pt, domain=0.2:8.0, smooth, samples=200, blue]
    plot (\x, {(\ymax-0.7)/(1+exp(-0.95*(\x-2.2))) + 0.3});

  \draw[dashed, purple!70, line width=3pt] (2.5,\ymin) -- (2.5,\ymax);
  \draw[dashed, purple!70, line width=3pt] (5.0,\ymin) -- (5.0,\ymax);
  \draw[<->, line width=2.5pt, purple!70] (2.5,1.5) -- (5.0,1.5); 
    \node[
        fill=green!15,
        rounded corners=4pt,
        inner xsep=2pt,
        inner ysep=4pt
    ] at (3.75,1.5) {$[N,2N]$};

\draw[->, thick, red!70] (2.8,2.8) -- (3.3,3.4)
    node[right, black, xshift=3pt, font=\LARGE] {$\dfrac{\mathrm{d\,Perf}}{\mathrm{d\log N}} \gg 0$};

\draw[->, thick, gray] (5.3,4.0) -- (5.7,4.0)   
    node[right, black, font=\LARGE] {$\mathrm{d\,Perf}/\mathrm{d\log N} \approx 0$};

\fill[blue] (2.5, {(\ymax-0.7)/(1+exp(-0.95*(2.5-2.2))) + 0.3}) circle (4pt)
    node[left, xshift=-6pt, yshift=8pt, font=\LARGE] {$\mathrm{Perf}_i(D_i)$};

\fill[blue] (5.0, {(\ymax-0.7)/(1+exp(-0.95*(5.0-2.2))) + 0.3}) circle (4pt)
    node[right, xshift=7pt, yshift=-9pt, font=\LARGE] {$\mathrm{Perf}_i(D_i + D_i')$};
    
  \node[anchor=center] at (1.25, 0.5) {Initial};
  \node[anchor=center, font=\LARGE\bfseries] at (3.75, 0.5) {Dynamic};
  \node[anchor=center] at (6.5, 0.5) {Saturation};
  \node[below, font=\LARGE] at (2.5,0) {$N$};
  \node[below, font=\LARGE] at (5.0,0) {$2N$};

\end{tikzpicture}}\vspace{-0.2cm}
\caption{Typical learning curve for a single language, showing the dynamic interval and derivative regimes.}
\label{fig:learning_curve_clean}
\end{figure}
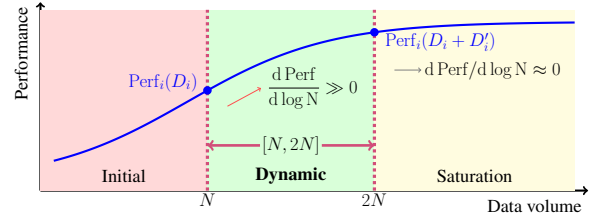\vspace{-0.35cm}

\begin{figure}[b]
\vspace{-0.2cm}
    \centering
    \begin{subfigure}[t]{0.48\linewidth}
        \vspace{0pt}
        \centering
        \includegraphics[width=\linewidth]{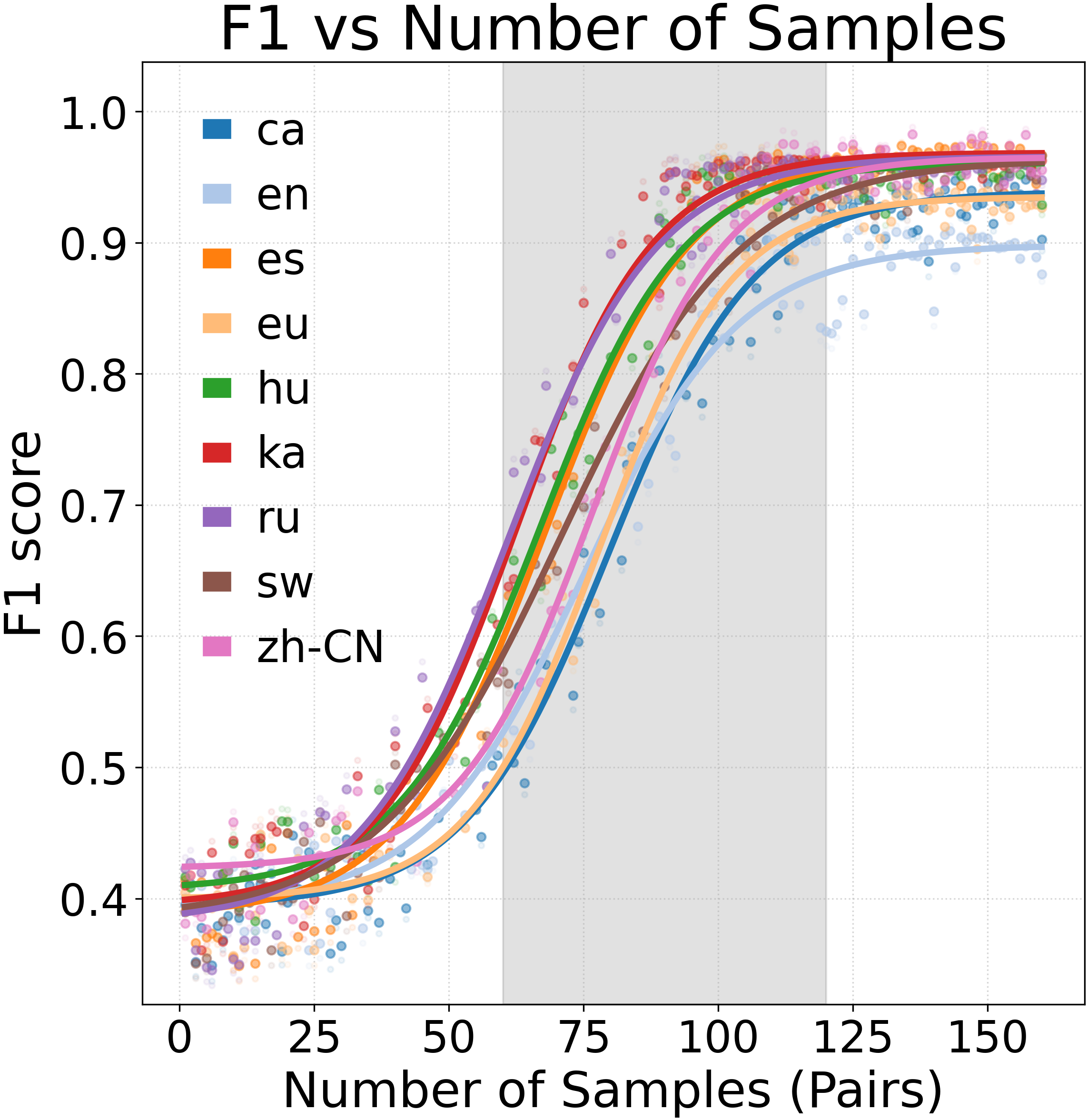}
        \caption{Gender Recognition}
        \label{fig:learning_curve_gender}
    \end{subfigure}
    \hfill
    \begin{subfigure}[t]{0.48\linewidth}
        \vspace{0pt}
        \centering
        \includegraphics[width=\linewidth]{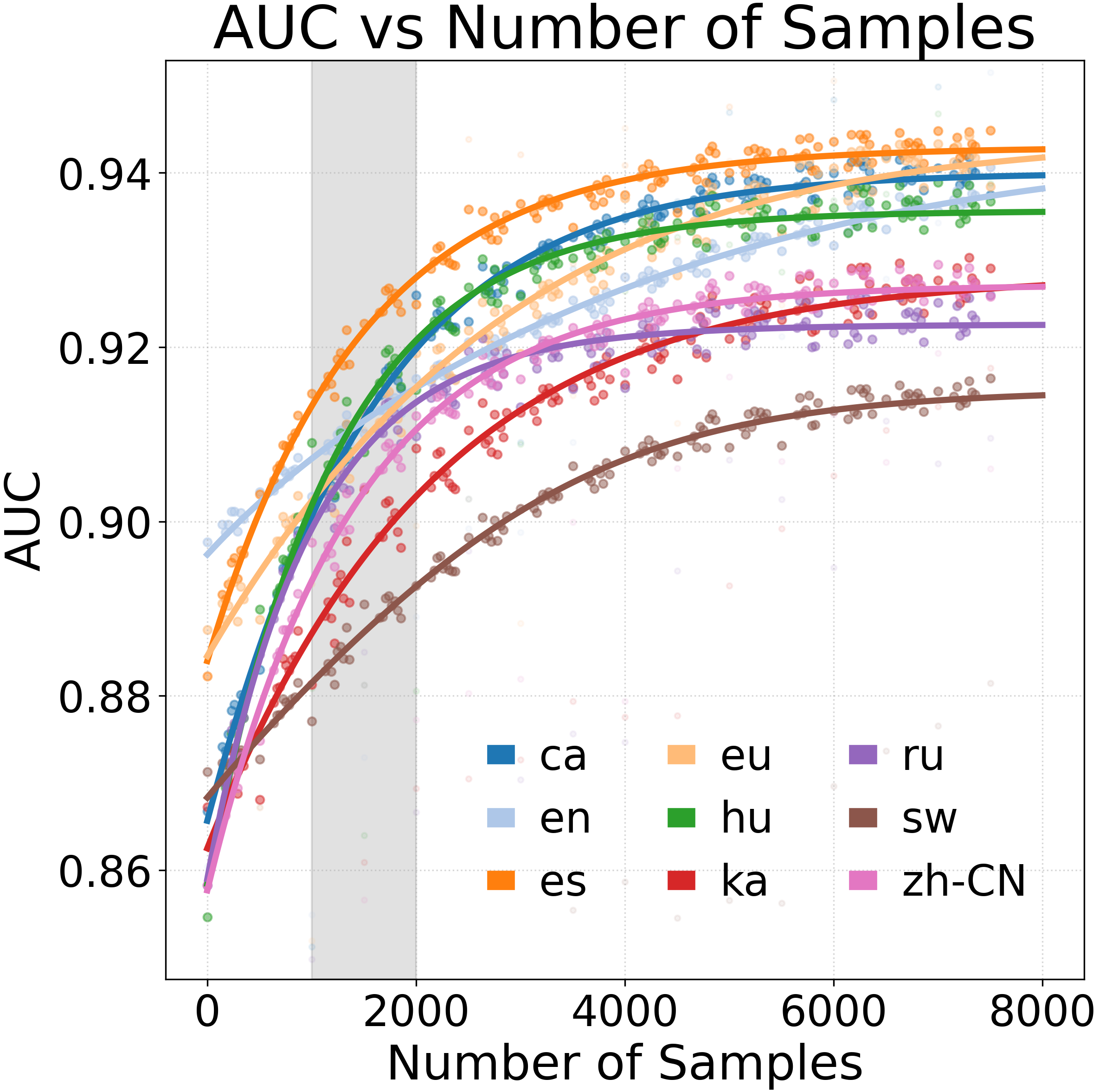}
        \caption{Speaker Verification}
        \label{fig:learning_curve_sv}
    \end{subfigure}\vspace{-0.2cm}
    \caption{Learning curves for both tasks, showing performance as a function of training samples for a representative subset of languages. The task-specific dynamic interval $[N,2N]$ used to compute the CLTM is highlighted.}
    \label{fig:learning_curves}
\end{figure}

\section{Experimental setup}
\label{sec:setup}
This section describes the experimental setup used to compute the CLTM. We evaluate the method on two paralinguistic tasks, Gender Recognition (GR) and Speaker Verification (SV), across 44 languages. Due to space constraints, only a representative subset of languages is shown in some visual analyses.

\subsection{Data}
\label{sec:data}

We use the Mozilla Common Voice corpus 22.0 \cite{ardila2020commonvoicemassivelymultilingualspeech} for its broad multilingual coverage and uniform data collection, which reduces extraneous variability. For each task, only samples with the required metadata are retained. Training data are strictly balanced across languages: equal numbers of samples and speakers (with language-disjoint speaker identities), a fixed number of 50 samples per speaker in SV, and balanced class distributions when applicable. This ensures that observed transfer effects are attributable to language rather than dataset composition. Evaluation uses the original test subset for each language, filtered only to retain samples with the required metadata. 

For each language $i$, a subset of $N$ samples is used to compute $\mathrm{Perf}_i(D_i)$ and augmented with a disjoint set of $N$ samples to compute $\mathrm{Perf}_i(D_i + D_i')$. Cross-lingual $\mathrm{Perf}_i(D_i + D_j)$ are obtained for all pairs $(i,j)$, $i \neq j$, by combining the corresponding $N$-sample subsets. 

The dynamic training interval (Section~\ref{sec:dat_regime}) was analyzed per task to ensure measurable self-gains ($\Delta_{i\leftarrow i}>0$) across all 44 target languages. Based on the learning curves (a representative subset shown in Figure~\ref{fig:learning_curves}), we select $[N,2N] = [60,120]$ pairs for gender recognition and $[N,2N] = [1000,2000]$ samples for speaker verification. All languages considered in this study satisfy $\Delta_{i\leftarrow i}>0$ within the task-specific interval.

\subsection{Model and Training}
\label{sec:model_training}

All experiments use the \texttt{mHuBERT-147}\footnote{Model checkpoint used in this work: \url{https://huggingface.co/utter-project/mHuBERT-147}} \cite{boito2024mhubert147compactmultilingualhubert} multilingual backbone, a HuBERT-based encoder pretrained on 147 languages (including the 44 analyzed here), ensuring a shared acoustic representation space across tasks and languages.

For downstream adaptation, a randomly initialized task-specific head is appended to the encoder. In both tasks, the head is a single linear classifier applied to the pooled encoder representation. The encoder is fine-tuned jointly with the head, with no frozen layers. Architectures and optimization are kept consistent across tasks so that observed differences reflect task characteristics rather than implementation choices.

Audio is resampled to 16 kHz and amplitude-normalized; all experiments use continuous (non-quantized) encoder representations. Training uses AdamW \cite{loshchilov2019decoupledweightdecayregularization} with a constant learning rate of $1\times10^{-5}$, weight decay $0$, gradient clipping (max norm $1.0$), and mixed-precision (fp16). Training is performed for a single epoch to minimize optimization-related confounds.

Because cross-lingual effects are often subtle, randomness is strictly controlled. Each reported result corresponds to the mean over $10$ independent seeds. Seeds are propagated to all relevant libraries and data-loading components to ensure deterministic initialization, shuffling, and batching.

\subsection{Downstream Tasks}
\label{sec:tasks}

\begin{figure}[t]
\centering
\includegraphics[width=0.95\linewidth]{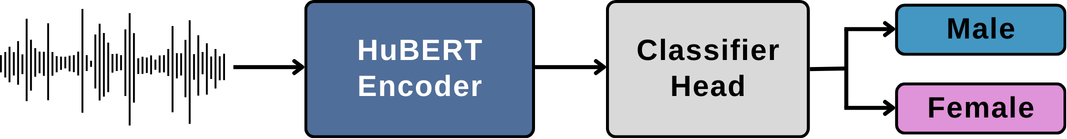}
\caption{Architecture for gender recognition.}
\label{fig:gender_pipeline}\vspace{-0.3cm}
\end{figure}

\begin{figure}[t]
\centering
\includegraphics[width=\linewidth]{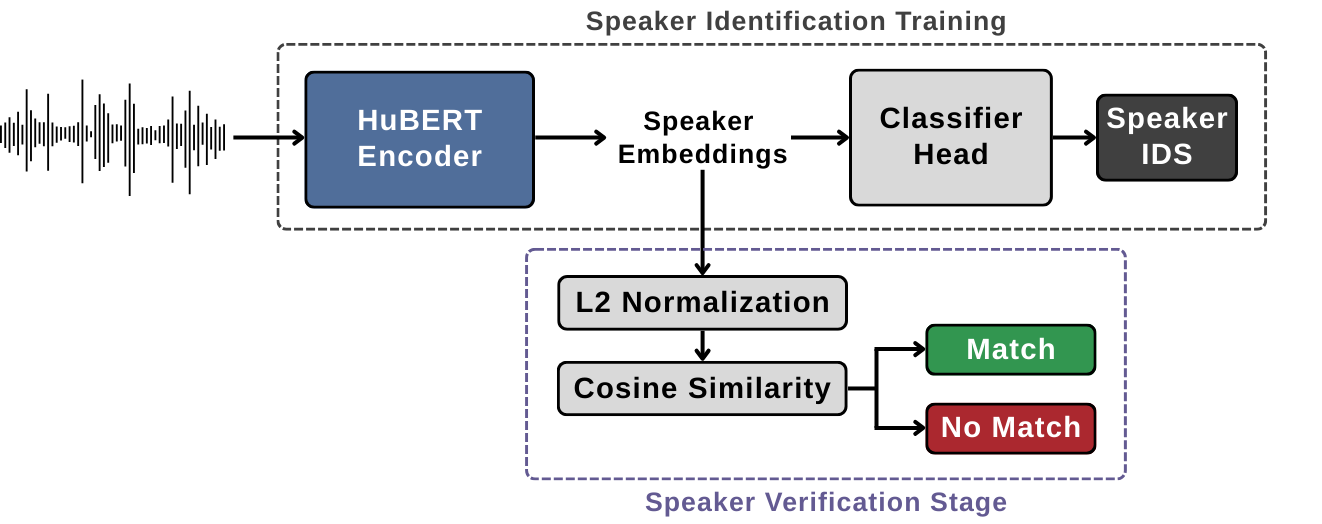}\vspace{-0.2cm}
\caption{Speaker verification pipeline: SID training via a classification head, then embeddings are L2-normalized and compared with cosine similarity for verification.}
\label{fig:sv_pipeline}\vspace{-0.3cm}
\end{figure}

\textbf{Gender Recognition}
Gender recognition is formulated as a binary classification task (male vs.\ female) with performance measured using macro-F1 to account for class imbalance. This task probes paralinguistic cues while remaining structurally simple; the overall architecture is illustrated in Figure~\ref{fig:gender_pipeline}.

\vspace{0.2cm}

\noindent\textbf{Speaker Verification}
Speaker verification assesses whether two utterances belong to the same speaker. 

We adopt a two-stage approach. First, a speaker identification (SID) classifier is trained to discriminate among training speakers. The classification head is then discarded and the fine-tuned backbone is used as a feature extractor (see Figure~\ref{fig:sv_pipeline}). Utterance-level embeddings are obtained by averaging the last hidden layers, applying temporal mean pooling, and L2-normalizing the resulting vectors. Similarity between embeddings is computed using cosine similarity. 

Evaluation follows a gender-controlled protocol, pairing negative samples by gender label to prevent reliance on this cue. Performance is reported as the Area Under the Curve (AUC).

\section{Results}
\label{sec:results}

For each task, we analyze the resulting matrix at two levels. First, we show reduced CLTMs for 16 representative languages as heatmaps using a symmetric color scale in $[-1.5,1.5]$, for a clear qualitative view of cross-lingual effects. 
The complete $44\times44$ matrices are available in the repository\footnote{The full $44\times44$ matrices and the code required to reproduce model training and all analyses are available at \url{https://github.com/Pol-Buitrago/cltm-framework}.}.

Second, we conduct an analysis of each matrix using the properties introduced in Section~\ref{sec:cltm}, providing a compact characterization of the geometry and underlying structure of cross-lingual interactions for each task.

\subsection{CLTM Qualitative Analysis}
\label{sec:qual}

\begin{figure*}[t]
\centering
\begin{subfigure}[b]{0.497\textwidth}
    \centering
    \includegraphics[width=\linewidth]{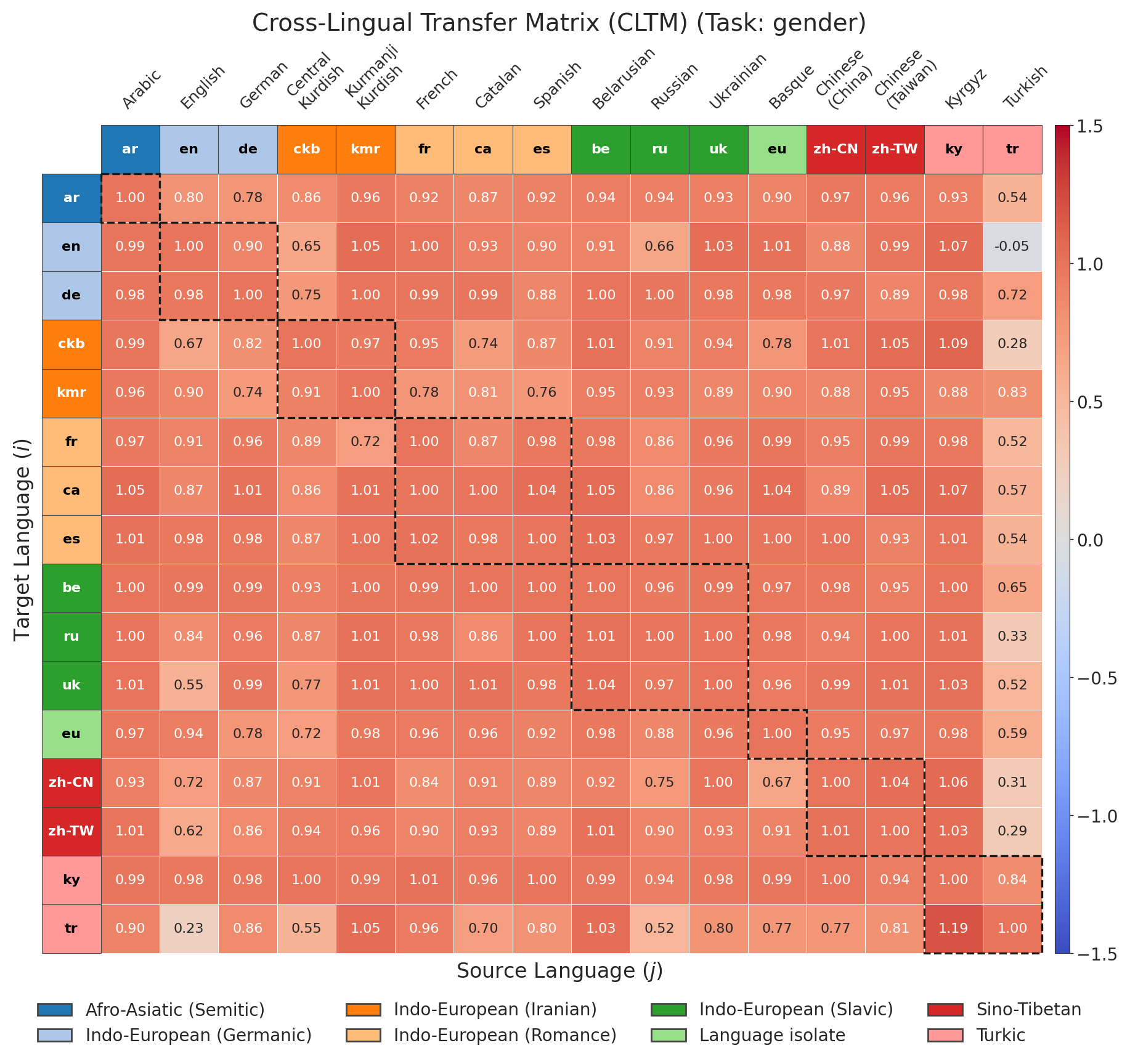}\vspace{-0.2cm}
    \caption{Gender Recognition}
    \label{fig:cltm_gender}
\end{subfigure}
\hfill
\begin{subfigure}[b]{0.497\textwidth}
    \centering
    \includegraphics[width=\linewidth]{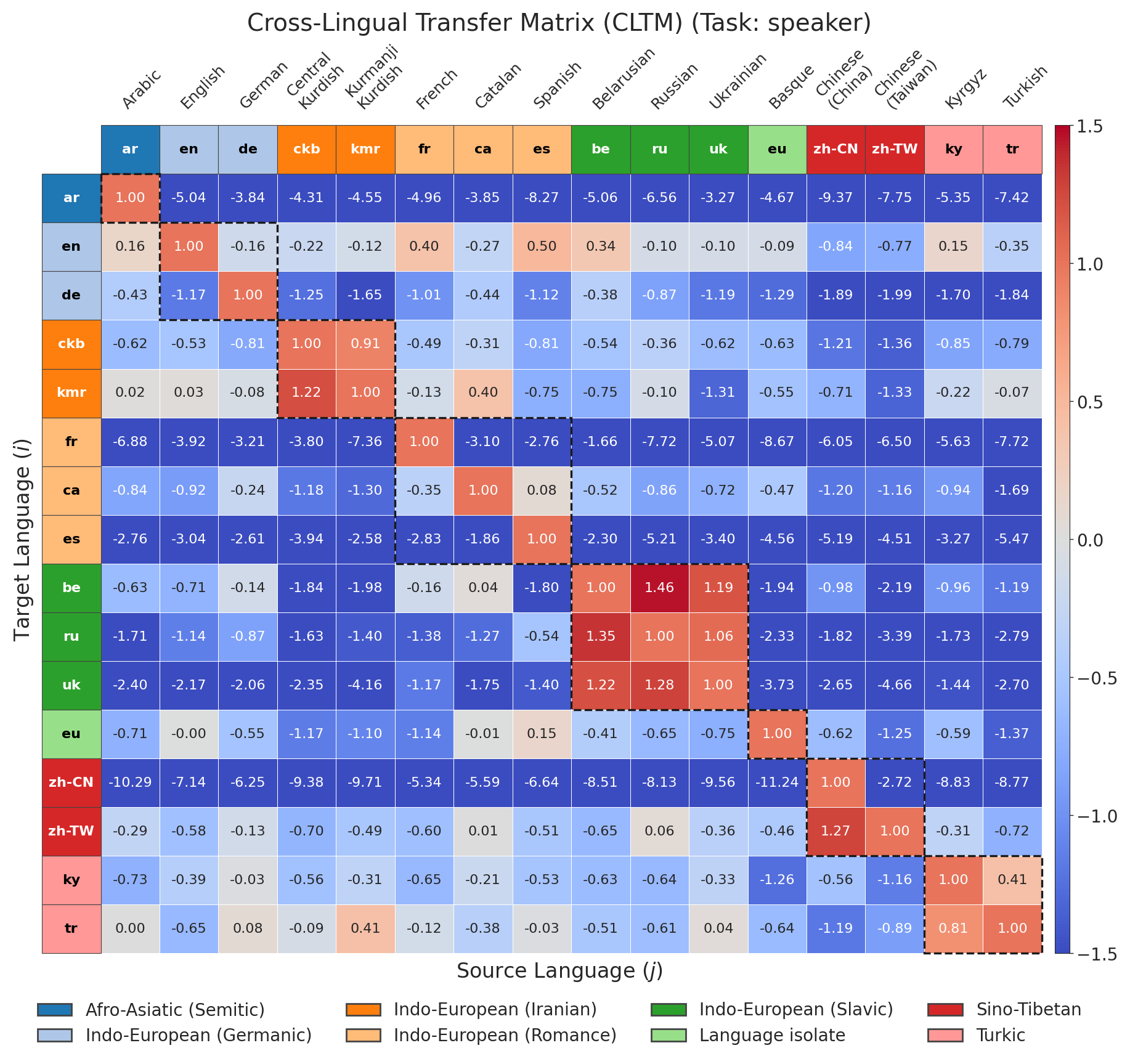}\vspace{-0.2cm}
    \caption{Speaker Verification}
    \label{fig:cltm_sv}
\end{subfigure}\vspace{-0.15cm}
\caption{Reduced CLTMs (16 representative languages) for gender recognition and speaker verification. Colors show how much adding donor-language data affects performance on a target language compared to adding the same amount of target-language data.}
\label{fig:cltm_all}\vspace{-0.2cm}
\end{figure*}

Figure~\ref{fig:cltm_all} shows reduced CLTMs for 16 representative languages, highlighting qualitative cross-lingual patterns. \vspace{0.2cm}

\noindent\textbf{Gender Recognition} For GR, the matrix is near the agnostic ideal, with most entries close to one and uniformly positive, showing largely language-independent transfer.\vspace{0.2cm}

\noindent\textbf{Speaker Verification} In contrast, SV shows strong language dependence. Negative transfer is widespread, whereas positive effects are sparse and often cluster near the diagonal, forming localized blocks within some language families. 
\vspace{0.1cm}

\subsection{CLTM Structural Analysis}

While CLTMs can be inspected visually to guide multilingual selection and identify qualitative patterns, their formulation also allows quantitative analysis through aggregate diagnostics. These metrics can be applied to any CLTM, regardless of task or architecture, to systematically compare cross-lingual transfer patterns and task-specific characteristics. 

Table~\ref{tab:cltm_diagnostics} summarizes these diagnostics for both tasks, computed on the full $44\times44$ CLTMs. The reported aggregate metrics corroborate the qualitative patterns observed in Section~\ref{sec:qual}.

\begin{table}[t]
\centering
\renewcommand{\arraystretch}{1.1} 
\caption{Aggregate CLTM diagnostics computed on the full $44\times44$ matrices for gender recognition and speaker verification.}
\vspace{-0.2cm}
\label{tab:cltm_diagnostics}
\resizebox{\columnwidth}{!}{%
\begin{tabular}{l @{\hspace{10pt}} c @{\hspace{20pt}} c}
\hline
\textbf{Metric} & \textbf{Gender Recognition} & \textbf{Speaker Verification} \\
\hline
$\mathrm{RFD}_1$ & 0.162 & 2.970 \\
$\mathrm{Asym}_{\mathrm{rel}}$ & 0.175 & 1.084 \\
$\text{prop}_{+}$ & 99.97\% & 8.93\% \\
$\text{reciprocity}_{+}$ & 99.93\% & 1.69\% \\
$\overline{\cos}_{\mathrm{rows}}$ & 0.990 & 0.615 \\
$\text{intra-family}_{+}$ & 4.98\% & 41.68\% \\
\hline
\end{tabular}%
}\vspace{-0.4cm}
\end{table}

\vspace{0.2cm}

\noindent\textbf{Gender Recognition} For GR, deviations from the language-agnostic ideal are minimal: Frobenius deviation and relative asymmetry are close to zero (with zero indicating perfect agnosticity and fully symmetric transfer). Donor-benefit profiles are highly consistent across targets, with $\overline{\cos}_{\mathrm{rows}}$ close to one (indicating nearly identical transfer profiles), and positive transfer is broadly distributed rather than concentrated within language families, confirming the absence of clear linguistic structure.\vspace{0.2cm}

\noindent\textbf{Speaker Verification} In contrast, SV shows strong language dependence. Frobenius deviation and asymmetry are substantially larger, reflecting heterogeneous transfer. Row similarity is lower, indicating distinct donor profiles across targets, and positive interactions ($\text{prop}_{+}$), though infrequent, concentrate within language families, showing that beneficial transfer occurs primarily among related languages.

\subsection{SV: Embedding Geometry}
\label{sec:sv_embeddings}

To explore potential sources of negative transfer in speaker verification, we report the Euclidean distances between language-specific centroids in the speaker embedding space ($d_{\mathrm{cent}}$) for two example pairs exhibiting positive transfer and two showing negative transfer (see Table~\ref{tab:sv_centroids_cltm}). Larger centroid distances appear associated with stronger negative transfer, suggesting that language-induced shifts in the embedding space may contribute to cross-lingual interference, potentially due to the SV architecture. Our exploratory experiments also indicate that multilingual models using this architecture tend to perform poorly overall, consistent with these observations.

\begin{table}[t]
\centering
\renewcommand{\arraystretch}{1.2} 
\caption{Euclidean centroid distances between language speaker embeddings and CLTM values for selected SV pairs.}\vspace{-0.2cm}
\resizebox{\linewidth}{!}{%
\begin{tabular}{lcc}
\hline
\textbf{Language pair} & $d_{\mathrm{cent}}$ (Euclidean) & CLTM values \\
\hline
Russian--Belarusian & 1.37 & 1.35 / 1.46\\
Kurmanji--Central Kurdish & 1.53 & 1.22 / 0.91 \\
German--Portuguese & 2.06 & -2.02 / -0.32 \\
Galician--Central Kurdish & 2.39 & -1.09 / -1.13 \\
\hline
\end{tabular}%
}
\label{tab:sv_centroids_cltm}\vspace{-0.4cm}
\end{table}

\subsection{Stability and Statistical Reliability of CLTM}
\label{sec:stability}

Because CLTM entries are ratios of performance differences, we assess their stability by computing both the median ($\tilde{\sigma}_\mathrm{SE}$) and the mean ($\bar{\sigma}_\mathrm{SE}$) standard error (SE) across the 10 seeds used. To contextualize the magnitude of the CLTM entries, we also report the root mean square ($\mathrm{RMS} = \lVert \mathrm{CLTM} \rVert_F / n$).

For gender recognition, seed variability is low relative to effect size ($\tilde{\sigma}_\mathrm{SE}=0.062$, $\bar{\sigma}_\mathrm{SE}=0.075$, $\mathrm{RMS}=0.935$). For speaker verification, variability is higher ($\tilde{\sigma}_\mathrm{SE}=0.468$, $\bar{\sigma}_\mathrm{SE}=0.545$) but so are effect magnitudes ($\mathrm{RMS}=2.25$). In both tasks, CLTM entries typically exceed SE, suggesting cross-lingual structure is not driven by seed variability.

Furthermore, all self-gains ($\Delta_{i\leftarrow i}$) are positive, with $\bar{\Delta}=0.304 \pm 0.089$ for gender recognition and $\bar{\Delta}=0.037 \pm 0.016$ for SV, confirming that the CLTM denominator is stable and that the resulting ratios reflect meaningful performance gains.

\section{Conclusions}
\label{sec:conclusions}

We introduce and validate the Cross-Lingual Transfer Matrix (CLTM), a normalized, performance-grounded measure of cross-lingual transfer. Using a HuBERT encoder with task-specific heads, CLTM reveals near language-agnostic transfer for gender recognition but strong language dependence for speaker verification, showing that paralinguistic tasks can still be sensitive to linguistic factors. CLTM provides interpretable and practical insights for multilingual data selection, while leaving open the exploration of additional tasks and architectures.

\newpage

\newpage
\section{Acknowledgements}


Project AIA2025-163322-C66 funded by MICIU/AEI/10.13039/501100011033. FC acknowledges his AI4S fellowship within the ``Generación D'' initiative by Red.es, Ministerio para la Transformación Digital y de la Función Pública, for talent attraction (C005/24-ED CV1), funded by NextGenerationEU through PRTR.

\section{Generative AI Use Disclosure}

The code, experiments, and manuscript presented in this work were developed and written by the authors. ChatGPT was used only in a limited manner for minor language polishing and occasional refinement of small code fragments. No generative AI tool was used to produce any substantial part of the manuscript, nor to generate or influence the methodology, experimental design, or scientific conclusions.

\bibliographystyle{IEEEtran}
\bibliography{mybib}

\end{document}